\documentclass{epl}

\title{Slow rolling, inflation, and quintessence}
\author{M. Capone\inst{1} \and C. Rubano\inst{2,3} \and P. Scudellaro\inst{2,3}}

\institute{
  \inst{1} Dipartimento di Fisica, Politecnico di Torino,
Corso Duca degli Abruzzi 24, I-10129 Torino, Italy\\
  \inst{2} Dipartimento di Scienze Fisiche, Universit\`a
Federico II di Napoli, Complesso Universitario di Monte S. Angelo,
Via Cintia, Ed. N, I-80126 Napoli, Italy\\
  \inst{3} Istituto Nazionale di Fisica Nucleare, Sez. Napoli,
Complesso Universitario di Monte S. Angelo, Via Cintia, Ed. G,
I-80126 Napoli, Italy}

\pacs{98.80.-k}{Cosmology}

\pacs{98.80.Jk}{Mathematical and relativistic aspects of
cosmology}

\pacs{98.80.Cq}{Particle-theory and field-theory models of the
early universe}

\begin{document}

\maketitle

\begin{abstract}
We comment on the choice of the quintessence potential, examining
the slow-roll approximation in a minimally coupled theory of
gravity. We make some considerations on the potential behaviors,
the related $\Gamma$ parameter, and their relationships to phantom
cosmology.
\end{abstract}

\section{Introduction}

In order to discriminate models for dark energy we usually assign
a scalar field potential $V =
V(\varphi)$~\cite{b.wetterich,b.peeblesratra,b.padma}. Within the
already available theoretical framework, the behaviors of its
first and second derivatives ($V^\prime \equiv dV/d\varphi$ and
$V^{\prime \prime} \equiv d^{2}V/d{\varphi}^2$) have been studied,
and some constraints from observational data have also been
derived. In this context, the $\Gamma$ function was first
introduced in~\cite{b.stein} to characterize acceleration properly
with a \textit{tracking} behavior, via a suitable theorem. Another
very popular (and closely related) approach is that of using the
so called \emph{scaling solutions}~\cite{b.ferrjoy1,b.ferrjoy2}.
As shown in~\cite{b.rspcc}, this first led to exclude possibly
good forms of the potential like the exponential one $V(\varphi)
\sim \exp(- \lambda \varphi)$, in fact shown in~\cite{b.rspcc} as
still deserving attention, since it meets all the usual
constraints posed by observational data. This was proved, in
particular, with $\lambda \equiv
\sqrt{3/2}$~\cite{b.rubanoscud,b.pavlov,b.rubanosereno,b.estersazhin,b.rspcc,b.demianski}.

Let us reconsider here some features of the potential for the
\emph{quintessence} $Q$ field~\cite{b.prlstein,b.stein}, also
thinking of the possibility of phantom cosmology, and comment on
the slow-rolling conditions introduced in early inflation.
Moreover, we consider the $\Gamma$ parameter and other aspects of
the potential with respect to the slow-roll approximation.

\section{Slow-roll conditions}

In inflation there is a phase of (almost) exponential cosmic
expansion. Its definition in terms of the scale factor $a$ of the
universe is given by a positive acceleration, $\ddot{a}>0$. In
standard cosmology this is a requirement on the cosmic content,
since if the $\Lambda$-term is zero or absorbed into the energy
density $\rho$ of a suitable fluid, we have $\rho +3p<0$
independently of the spacetime curvature. Being $\rho$ positive,
we must introduce a fluid with negative pressure, $p<0$, for
example a single scalar field $\varphi$. Assuming homogeneity and
isotropy, the energy density $\rho_{\varphi}$ and pressure
$p_{\varphi}$ are written in terms of \emph{kinetic} and
\emph{potential} contributions
\begin{equation}
\rho_{\varphi} = \frac{1}{2}\dot{\varphi}^{2}+V\left( \varphi
\right) \,,\,\,\,\, p_{\varphi} =
\frac{1}{2}\dot{\varphi}^{2}-V\left( \varphi \right) \,,
\label{e.3}
\end{equation}
where an inflationary model is given once we assume a potential
$V$.

The Friedman equations of a homogeneous and isotropic spatially
flat universe are ($c = 1$)
\begin{equation}
H^{2} = \frac{8\pi G}{3}\left( \rho _{m}+V\left( \varphi \right)
+\frac{1}{2}\dot{\varphi}^{2}\right) \,,  \label{e.5}
\end{equation}
\begin{equation}
\dot{H} = -4\pi G\left( \rho _{m}+p_{m}+\dot{\varphi}^{2}\right)
\,, \label{e.6}
\end{equation}
and
\begin{equation}
\ddot{\varphi}+3H\dot{\varphi}+\frac{dV}{d\varphi}=0 \,,
\label{e.7}
\end{equation}
where $H\equiv \dot{a}/a$, and $\rho _{m}$ and $p_{m}$ are the
energy density and pressure of the matter component, which are
null in early inflation. In this last case, the \emph{slow-roll
approximation}~\cite{b.lpb,b.liddlelith} leads to discard the last
term in eq.~(\ref{e.5}) and the first term in eq.~(\ref{e.7}), so
that
\begin{equation}
H^{2} \simeq \frac{8\pi G}{3}V \,,\,\,\,\, \dot{H} \simeq -4\pi
G\dot{\varphi}^{2} \,,\,\,\,\, 3H\dot{\phi}\simeq -V^{\prime} \,.
\label{e.8}
\end{equation}
This is equivalent to requiring that: i) the universe is in a
phase of acceleration; ii) the field is slowly varying, or slowly
rolling down its potential ($\dot{\phi}^{2}\ll 2V$); iii) the
acceleration of the scalar field is also small ($\ddot{\varphi}
\ll 3H\dot{\varphi}$). This is sufficient to guarantee inflation,
but it is not also necessary, since in principle inflation can
take place even if slow-roll conditions are
violated~\cite{b.liddlelith}.

Let us introduce the two \textit{slow-roll
parameters}~\cite{b.lpb,b.liddlelith}
\begin{equation}
\varepsilon \left( \varphi \right) \equiv \frac{3}{2}\left(
1+w_{\varphi }\right) \,,\,\,\,\, \eta \left( \varphi \right)
\equiv \frac{\ddot{\varphi}}{H\dot{\varphi}} \,, \label{e.11}
\end{equation}
where $w_{\varphi }\equiv p_{\varphi }/\rho _{\varphi }$ is the
equation of state of the scalar fluid, and require that
\begin{equation}
\varepsilon \left( \varphi \right) \ll 1\,,\,\,\,\, \left\vert
\eta \left( \varphi \right) \right\vert \ll 1 \,. \label{e.13}
\end{equation}
In scalar field cosmology, when ${\dot{\varphi}}^2 = - 2\dot{H}$,
$\varepsilon$ and $\eta$ are related to $H$ and its time
derivatives
\begin{equation}
\varepsilon = - \frac{\dot{H}}{3H^2 + \dot{H}}\,, \,\,\,\,\,\,\,
\eta = \frac{\ddot{H}}{2H\dot{H}} \,. \label{e.13-1}
\end{equation}

Anyway, we think it is important to discuss the slow-roll
approximation also when matter is present. Thus, we go back to
eq.~(\ref{e.7}) and differentiate it with respect to time, and,
from eq.~(\ref{e.6}) (with $p_{m} = 0$), get
\begin{equation}
V^{\prime \prime} \dot{\varphi} = 12\pi G\dot{\varphi}(\rho_m +
{\dot{\varphi}}^2) - 3 H^2 \eta \dot{\varphi} -
\frac{d(\ddot{\varphi})}{dt}\,. \label{e.13-2}
\end{equation}
>From $\dot{\varphi} \neq 0, V \neq 0$ and $d(\ddot{\varphi})/dt
\ll \dot{\varphi}$, we thus find (for $\varepsilon \neq 3$)
\begin{equation}
\frac{V^{\prime \prime}}{V} = 4\pi G\left[
(3-2\eta)\frac{\rho_m}{V} +
\frac{6(\varepsilon-\eta)}{3-\varepsilon} \right]\,.
\label{e.13-3}
\end{equation}
(The value $\varepsilon = 3$ is excluded here and means
$w_{\varphi} = 1$, typical for stiff matter.) We can also rewrite
eq.~(\ref{e.7}) as $(V^{\prime})^2 = H^2 {\dot{\varphi}}^2 (3 +
\eta)^2$ and get
\begin{equation}
\left( \frac{V^{\prime}}{V} \right)^2 = \frac{16\pi
G\varepsilon}{3}\left( \frac{3 + \eta}{3 - \varepsilon} \right)^2
\left[ 3 + (3 - \varepsilon)\frac{\rho_m}{V} \right]\,.
\label{e.13-5}
\end{equation}
We have now two subcases we discuss separately in the
following.

\subsection{Inflation}

With $\rho_m =0$, if $\varepsilon \ll 1$ we can Taylor expand
$w_{\varphi}$ to the first order in $\dot{\varphi}^{2}/(2V)$, and
obtain $w_{\varphi }\simeq -1+\dot{\varphi}^{2}/V$, while, due to
the first of eq.~(\ref{e.11}), we find $\varepsilon \simeq
3\dot{\varphi}^{2}/(2V)$. With $\eta \ll 1$,  eq.~(\ref{e.7})
tells us that $\dot{\varphi}^{2}\simeq V^{\prime 2}/(16H^{2})$,
and eq.~(\ref{e.5}), with $\varepsilon \ll 1$, becomes the first
of eq.~(\ref{e.8}). Assuming eq.~(\ref{e.13}), it is not difficult
to prove that, to the first order, it is
\begin{equation}
\varepsilon \simeq \frac{1}{16\pi G}\left( \frac{V^{\prime
}}{V}\right)^{2} \,,\,\,\,\, \varepsilon -\eta \simeq
\frac{1}{8\pi G}\frac{V^{\prime \prime }}{V} \,. \label{e.17}
\end{equation}
Thus, we can find that, if the universe is in the slow-rolling
regime, both the relative slope and curvature of the potential $V$
are small, being $V^{\prime }/V\ll 1\left( GeV\right) ^{-1}$ and
$V^{\prime \prime }/V\ll 1\,\left( GeV\right)^{-2}$. (With $\hbar
=c=1$, $G=6.842\times 10^{-39}\left( GeV\right)^{-2}$, and $\left[
\varphi \right] =eV,$ $\left[ \dot{\varphi}^{2}\right] =\left[
V\right] =\left[ \rho _{\varphi }\right] =eV^{4}$, $\left[
\ddot{\varphi}\right] =\left[ V^{\prime }\right] =eV^{3}$, $\left[
V^{\prime \prime }\right] =eV^{2}$.)

The relations in eq.~(\ref{e.17}) are found \emph{assuming}
$\varepsilon, \eta \ll 1$, so that they cannot be used to prove
that small $V^{\prime}/V$ and $V^{\prime \prime}/V$ imply
$\varepsilon, \eta$ to be small, too. To show this, let us rewrite
eq.~(\ref{e.13-5}) (with $\rho_m = 0\,, V^{\prime}/V \ll 1\,\left(
GeV\right)^{-1}\,, V^{\prime \prime}/V \ll 1\,\left(
GeV\right)^{-2}$, and $d(\ddot{\varphi})/{dt} \ll \dot{\varphi}$)
and the first of eq.~(\ref{e.11})
\begin{equation}
V^{\prime \prime }\simeq 12\pi G\dot{\varphi}^{2}-3H^{2}\eta
\,,\,\,\,\, \dot{\varphi}^{2}=\frac{2\varepsilon V}{3-\varepsilon}
\,, \label{e.20}
\end{equation}
so that from eqs.~(\ref{e.6}) and the first of (\ref{e.20}) we get
\begin{equation}
\frac{V^{\prime \prime }}{V}\simeq 24\pi G(\varepsilon -\eta) \,.
\label{e.22}
\end{equation}

Also, the scalar field equation can be written as $V^{\prime
2}=H^{2}\varphi ^{2}\left( 3+\eta \right) ^{2}$, and from both the
first Friedman equation and the second of eq.~(\ref{e.20}) it is
\begin{equation}
\left( \frac{V^{\prime }}{V}\right) ^{2}\simeq 16\pi G\varepsilon
\left( \frac{3+\eta }{3-\varepsilon }\right)^{2} \simeq 16\pi
G\varepsilon \,. \label{e.24}
\end{equation}
So, we obtain the two equations (\ref{e.22}) and (\ref{e.24}) for
$\varepsilon$ and $\eta$. To the first order in $V^{\prime }/V$
and $V^{\prime \prime }/V$, they give eq.~(\ref{e.17}) again.

Thus, in the early inflationary scenario asking for $\varepsilon$
and $\eta$ to be small is a necessary and sufficient condition for
$V^{\prime }/V$ and $V^{\prime \prime}/V$ to be small, too. The
inflaton potentials are just selected through their relative
slopes and curvatures.

\subsection{Quintessence}

Now, there is no fundamental reason why the scalar field should
obey to the slow-roll approximation. There is a difficulty of
relating the slow-roll parameters with the potential today, which
is mainly due to the impossibility of neglecting the ordinary
matter contribution. Let us in fact consider $V = V(Q)$ and, as
supposed before, assume that $\dot{Q}\neq 0$ and also that
$d(\ddot{Q})/{dt}$ is negligible with respect to $Q$ and
$\dot{Q}$. From eqs.~(\ref{e.13-3}) and (\ref{e.13-5}), imposing
$\varepsilon ,\eta \ll 1$ does not imply that the relative slope
and curvature be small again. The same is true for the viceversa.
To make it possible, $\rho_{m}/V$ should in fact be at least of
the same order of magnitude of $\varepsilon$ and $\eta$, but now
the ratio $\rho_{m}/V$ is of order unity. Even if today $Q$ is
such that ${\dot{Q}}^2<2V(Q)$, so giving rise to acceleration, it
is also not necessarily \emph{slowed down}. Thus, $\varepsilon$
and $\eta$ are now not so crucial as before in the cosmological
description, while keeping information on the local behavior of
the scalar field.

As an example, consider the dark energy model (with both $Q$ and
CDM) studied in~\cite{b.rubanoscud}, with an exponential potential
$V(Q)\sim \exp (- \sqrt{3/2} Q)$ ($8\pi G = 1$); the equations are
generally and exactly solved, leading to~\cite{b.rubanoscud}
\begin{equation}
w_Q = - \frac{3 + 2{\tau}^2}{3 + 4{\tau}^2}\,,\,\,\,\, H =
\frac{2(1 + 2{\tau}^2)}{3t_s \tau(1 + {\tau}^2)}\,, \label{e.24-1}
\end{equation}
where $t_s$ is a time scale such that $H(t_s) \equiv {t_s}^{- 1}$,
i.e., of the order of the age of the universe, and $\tau \equiv
t/t_s = H(t_s)\,t$ a suitable dimensionless time taking the value
$\tau \simeq 1$. Posing $t_s = 1$ for simplicity, from
eq.~(\ref{e.13-1})
\begin{equation}
\varepsilon = - 1 + \frac{2(1 + 2{\tau}^2)^2}{1 + 7{\tau}^2 +
6{\tau}^4}\,,\,\,\,\, \eta = - \frac{3(1 + 3{\tau}^2 +
2{\tau}^6)}{2(1 + 3{\tau}^2 + 4{\tau}^4 + 4{\tau}^6)}\,.
\label{e.24-3}
\end{equation}
So, choosing for instance $\tau = 0.82$ (as
in~\cite{b.rubanoscud}) leads to $\varepsilon \simeq 0.31$ and
$\eta \simeq -0.90$, which is not consistent with slow roll (i.e.,
$\varepsilon, \eta \ll 1$).

We can also see that, on the other hand, if we generally assume
\begin{equation}
V(Q) = V_0 \exp \left( - \sqrt{\frac{2}{p}} \right)Q\,,\,\,\,\,
a(t) = a_0 t^p\,, \label{e.41-1}
\end{equation}
with $p$ a generic parameter, we find
\begin{equation}
Q(t) = \sqrt{2p}\ln \left( \sqrt{\frac{V_0}{p(3p - 1)}}t
\right)\,. \label{e.41-3}
\end{equation}
This means that we have
\begin{equation}
H(t) = \frac{p}{t}\,,\,\,\,\, \dot{Q} =
\sqrt{\frac{2p}{t}}\,,\,\,\,\, \ddot{Q} = -
\frac{\sqrt{2p}}{t^2}\,,\,\,\,\, V(t) = p(3p -1)t^{-2}\,,
\label{e.41-4}
\end{equation}
satisfying eqs.~(\ref{e.7}) and (\ref{e.5}). On the other hand,
since in general
\begin{equation}
\varepsilon = \frac{3{\dot{Q}}^2}{2V} = \frac{3}{3p -
1}\,,\,\,\,\, \eta = \frac{\ddot{Q}}{H\dot{Q}} = - \frac{1}{p}\,,
\label{e.41-6}
\end{equation}
$p = 4/3$ (and $2/p = 3/2$) in fact implies $\varepsilon = 1$ and
$\eta = - 3/4$, so that such parameters are not negligible with
respect to $1$.

Furthermore, $\varepsilon$ needs to be greater than zero for the
\emph{usual} accelerated universe. As a matter of fact, when
$\varepsilon < 0$ (and, as usual, $V>0$), we get ${\dot{Q}}^2<0$
from eq.~(\ref{e.41-6})$_{1}$. That is, we are in presence of
\emph{phantom cosmology}~\cite{b.caldwell}, with an equation of
state $w_Q < - 1$ and the prediction of a \emph{big rip} in the
future~\cite{b.ckw}. Although the kinetic energy of $Q$ has to be
negative~\cite{b.cht}, this kind of cosmology is indeed compatible
with observational constraints, such that $- 1.38 < w_Q < -
0.82$~\cite{b.melchiorri}. A related interesting example can be
found in~\cite{b.nmc-rs}, where exact solutions for scalar-tensor
theories are used to implement dark energy models with varying $G$
and $\Lambda$, such that phantom cosmology can be recovered
without any big rip, since $\rho_Q$ in fact results always
decreasing in the future. (This appears to be peculiar, indeed, to
scalar fields nonminimally coupled to gravity.)

Now, let us discuss what happens when $V^{\prime \prime}/V$ and
$V^{\prime}/V$ are small. In order to understand whether and when
eqs.~(\ref{e.13-3}) and (\ref{e.13-5}) lead to contemporary small
values for $V^{\prime \prime}/V$ and $V^{\prime}/V$, we here
choose to present only indicative reasonings. For sake of
simplicity, let us in fact assume that $V^{\prime \prime}/V$ and
$V^{\prime}/V$ are both so small that we can take them directly
null
\begin{equation}
\frac{V^{\prime \prime}}{V} = 0 = (3 - 2\eta)\delta +
\frac{6(\varepsilon - \eta)}{3 - \varepsilon} \label{e.25}
\end{equation}
and
\begin{equation}
\left( \frac{V^{\prime}}{V} \right)^2 = 0 = \varepsilon \left(
\frac{3 + \eta}{3 - \varepsilon} \right)^2 \left[ 3 + (3 -
\varepsilon)\delta \right]\,, \label{e.26}
\end{equation}
where we posed $\delta \equiv \rho_m/V > 0$ (always, for $V>0$).
Solving for $\varepsilon$ and $\eta$ gives
\begin{equation}
\varepsilon = 3\frac{3\delta + 2}{3\delta - 2}\,,\,\,\,\, \eta =
-3 \,, \label{e.26-1}
\end{equation}
which shows that $\varepsilon < 0$ (corresponding to phantom
cosmology) when $\delta < 2/3$. In this approximated situation, we
can examine better which are the allowed non contradictory values
for $\varepsilon$ and $\eta$, so finding that, with the assumption
$\delta > 0$ (including also phantom cosmology), we can only
accept couples of values of $\varepsilon, \eta$ such that
\emph{either} $\varepsilon \neq 0, \eta = - 3$ \emph{or}
$\varepsilon = 0, \eta \neq \pm 3$. Anyway, we must not forget
that we have taken $V^{\prime \prime}/V = V^{\prime}/V = 0$, that
is, a more extreme case than the one with $V^{\prime \prime}/V$
and $V^{\prime}/V$ simply small. But we have also seen that our
hypotheses cannot yield both $\varepsilon = 0$ and $\eta = 0$,
since eq.~(\ref{e.25}) in fact becomes $\delta = 0$. This then
implies that the usual slow-rolling regime only belongs to a
scalar-field dominated universe, typical for early inflation but
not for quintessence today.

As a final side remark, note that, last but not least, assuming
$\varepsilon = 0$ is equivalent to consider $w_Q = - 1$, i.e., the
cosmological constant.

\section{The $\Gamma$ function}

The $\Gamma$ function is defined as
\begin{equation}
\Gamma \equiv \frac{VV^{\prime \prime }}{\left( V^{\prime }\right)
^{2}} = \frac{V^{\prime \prime}/V}{(V^{\prime}/V)^2} \,,
\label{e.27}
\end{equation}
and was first introduced in~\cite{b.stein}. It then revealed not
so interesting as supposed~\cite{b.bludmanroos, b.rspcc}, since it
is not necessary now to consider slow roll for the $Q$-field.
Anyway, $\Gamma$ still remains a significant indicator, since it
is skillfully built from a suitable ratio with the derivatives of
the potential. Even if it has been shown that $\Gamma > 1$ cannot
be considered as obvious for a correct quintessence tracking
behavior, that remains a sufficient condition~\cite{b.rspcc}. (See
also~\cite{b.bludmanroos} for related illuminating comments.) For
example, a single exponential ($\Gamma = 1$), or a suitable
combination of two of them ($\Gamma < 1$), is indeed compatible
with observational constraints~\cite{b.rspcc,b.demianski}.

It is interesting to find all the potentials giving rise to
strictly constant values of $\Gamma$. Considering eq.~(\ref{e.27})
as a differential equation for $V = V(Q)$, a direct investigation
in fact gives (for a constant $\Gamma$):

i) $\,\,\,\,\Gamma = 0 \Rightarrow V = \alpha Q + \beta$,

ii) $\,\,\Gamma = 1 \Rightarrow V = \beta \exp(-\alpha Q)$,

iii) $\Gamma = -1 \Rightarrow V = \sqrt{\alpha Q + \beta}$,

iv) $\,\,\Gamma \neq 0,\pm 1 \Rightarrow V = (\alpha Q +
\beta)^{1/(1 - \Gamma)}$, \\
where $\alpha, \beta$ are suitable integration constants. Case
iii) includes the well known and most commonly used potential $V
\sim Q^{-\lambda}$ (for $\Gamma =1+1/\lambda$).

When $w_Q$ is nearly constant, with $w_{m} \approx 0$ (i.e., an
asymptotic dominance of the scalar field), we anyway have $\Gamma
\approx 1$. In such a case, the potential must be very close to
exponential (see also~\cite{b.rubbarrow}). But the condition on
$w_Q$ is strictly verified only asymptotically, so indicating that
the potential must be exponential there, whatever its functional
form before. (Note that an exponential potential in fact leads to
some variation of the values of $w_Q$ in the present
period~\cite{b.rspcc}.)

Assuming both the conditions in eq.~(\ref{e.13}) and $V = V_0
\exp(- \lambda \varphi)$ (with $\lambda$ a generic constant),
eq.~(\ref{e.17}) now transforms into
\begin{equation}
\varepsilon \simeq \frac{1}{16 \pi G} {\lambda}^2 \,,\,\,\,\,
\varepsilon - \eta \simeq \frac{1}{8 \pi G} {\lambda}^2\,,
\label{e.33}
\end{equation}
which evidently lead to $0 < \varepsilon \simeq - \eta \ll 1$. If
we instead assume $V = V_0 {\varphi}^{- \alpha}$ ($\alpha
> 0$), eq.~(\ref{e.17}) gives
\begin{equation}
\varepsilon \simeq \frac{1}{16 \pi G}
\frac{{\alpha}^2}{{\varphi}^2} \,,\,\,\,\, \varepsilon - \eta
\simeq \frac{\alpha(1 + \alpha)}{8 \pi G {\varphi}^2} \,,
\label{e.36}
\end{equation}
so that $0 < \varepsilon \simeq \alpha \eta /(2 + \alpha)$.

We have used $\varphi$ instead of $Q$ since this is valid only in
the early inflationary period; from the definition of $\Gamma$,
and using eq.~(\ref{e.17}) (as well as eqs.~(\ref{e.22}) and
(\ref{e.24})), we in fact find
\begin{equation}
\Gamma \equiv \frac{VV^{\prime \prime }}{\left( V^{\prime }\right)
^{2}} \simeq \frac{\varepsilon - \eta}{2 \varepsilon}
\,,\label{e.39}
\end{equation}
which of course reproduces both the results found above for
$\varepsilon$ in the early inflationary stage.

Also, eqs.~(\ref{e.13-3}) and (\ref{e.13-5}) lead to the general
expression
\begin{equation}
\Gamma \equiv \frac{VV^{\prime \prime }}{\left( V^{\prime }\right)
^{2}} = \frac{3}{4\varepsilon}\frac{(3 - 2\eta)(3 - \varepsilon)^2
\delta + 6(\varepsilon - \eta)(3 - \varepsilon)}{(3 + \eta)^2 ([3
+ \delta (3 - \varepsilon)]} \simeq \frac{2}{3}\frac{V - \rho_m}{V
+ \rho_m} + \frac{1}{\varepsilon}\left( \frac{\rho_m}{V + \rho_m}
+ \frac{2}{3}\eta \right) \,,\label{e.40}
\end{equation}
where $\varepsilon \neq 0$ is assumed. When $\delta = 0$
(asymptotic scalar field dominance), this gives back
eq.~(\ref{e.39}) for $\varepsilon, \eta \ll 1$; when $\delta \neq
0$, eq.~(\ref{e.40}) implies (also without taking $\varepsilon,
\eta \ll 1$)
\begin{equation}
\Gamma \simeq \frac{3\delta + 2(\varepsilon - \eta)}{4\varepsilon
(1 + \delta)} \simeq \frac{3\delta}{4\varepsilon (1 + \delta)}
\label{e.40-3}
\end{equation}
for $\delta \simeq 1$. This is equivalent to $\Gamma \gg 1$ but is
not always valid, due to the possible asymptotic behavior $\delta
\rightarrow 0$, when the $Q$ content completely dominates the
universe. It can be characteristic, together with
eq.~(\ref{e.39}), only for cosmology with $w_Q > - 1$.

It is interesting to notice that we recover a phantom energy
scenario (with $w_Q < - 1$) only for $\Gamma < 0$, as can be soon
seen from eqs.~(\ref{e.41-6}) and (\ref{e.40-3}). This sheds new
light on how the parameter $\Gamma$ works with this kind of
energy.

The presence of ordinary matter today complicates the relationship
between $\varepsilon$ and $\eta$. We have already seen that the
condition $\varepsilon \ll 1$ is in fact tuned by the non
negligible ratio $\delta \equiv \rho_m/V(Q)$. This is well
illustrated, for example, assuming again an exponential potential.
In this case, eq.~(\ref{e.40}) gives
\begin{equation}
\varepsilon \simeq - \frac{1}{1 + 5\delta}[3 \delta + 2(1 +
\delta)\eta] \,, \label{e.41}
\end{equation}
while eq.~(\ref{e.40-3}), valid for $\delta \neq 0$ and
$\varepsilon, \eta \ll 1$, yields
\begin{equation}
\varepsilon \sim \frac{-2\eta + 3\delta}{3(1 + 2\delta)}\,.
\label{e.41-0}
\end{equation}
So, we find that $\varepsilon < 0$ (phantom cosmology) only when
$\eta > 3\delta/2 > 0$, being $\delta > 0$ always. This then
inverts the sign contraposition between $\varepsilon$ and $\eta$
found for the exponential potential in the early inflationary
stage.

>From the above considerations it appears, first of all, that the
situation usually depicted in quintessence is in fact quite
different from the inflationary scenario. This seems a rather
trivial observation, but, as a matter of fact, some confusion or
lack of clearness on this point is still present in the
literature.

The main results of this paper lie in a careful analysis of the
relationship between slow rolling parameters and the $\Gamma$
function. This goes in the direction to illustrate the problems
related to slow roll, also illuminating how the phantom energy may
enter the game. Such considerations are actually interesting for
the remarkable role this strange kind of energy is playing in
cosmology recently. Only new data will give a way to discriminate
in the near future among the various proposals till now posed by
theory, but now it still makes sense to speculate on them.

\end{document}